\documentclass[11pt]{article}
\usepackage{latexsym}
\usepackage{amsfonts}
\usepackage{amssymb}
\usepackage{amsmath}
\usepackage{url}

\begin{document}

\title{A Comment on Budach's Mouse-in-an-Octant Problem}
\author{Amir M. Ben-Amram\thanks{\texttt{amirben@mta.ac.il}}}

\maketitle

The decidability of Budach's Mouse-in-an-Octant Problem~\cite{vEB:1980,pultr1982} is apparently still open.
This note sketches a proof that an extended version of the problem (a super-mouse) is undecidable.

\subsection*{The Problem}

In the original formulation of the problem, the Budach mouse is a kind of finite automaton with
a very simple program, a cycle of states, $q_0 \xrightarrow{d_1} q_1 \xrightarrow{d_2} 
\dots q_{n-1} \xrightarrow{d_n} q_0$, with $d_1,\dots,d_n \in \{N,E\}$.
The mouse moves among integer points in the octant of the Cartesian plane delimited by the lines
$y=0$ and $y=x$. It begins on $(1,0)$, at state $q_0$, and advances in steps which increase the
state number (cyclically) and move the mouse, one unit at a time, in the direction indicated in the program
(North or East). If the mouse hits the line $y=x$, its position is reset from $(x,x)$ to $(x,0)$, without
affecting the control state. Operation proceeds as usual. The computational problem is to decide,
given the mouse's program and a distinguished state number $k$, whether the mouse will ever visit
the line $y=x$ while at state $q_k$.  We may refer to this event as a stopping condition, giving 
the problem the flavour of a halting problem.
Whether this problem is decidable is not known.

In this note, we increase the power of the
mouse by allowing several cycles, with states denoted $q_{c,0},\dots,q_{c,n_c-1}$ where $c$ is
the index of the cycle. A function $\delta$ maps states to cycles, and its use is that if
the mouse hits the line $y=x$ in state $q_{c,i}$, not only is it returned to the $x$ axis, but also its
state is set to the initial state of the cycle $\delta(q_{c,i})$.  

We shall prove that the decision problem for this mouse is undecidable, by reducing from the problem: does a given
2CM halt, when started with null counters?

\subsection*{2-Counter Machines}

A machine is specified by a list of $N$
instructions. Instruction $s$. where
$0\le s< N$ is the instruction number (can be seen as the control state of the machine),
is a pair $(D_1,D_2)$, such that $D_j$ is either $+1$, $-1$ or $0$ and represents a requested change to
$R_j$, followed by
 a list of 3-tuples of the form $(b_1, b_2,$ $k)$,
where $b_j \in \{Z, P\}$ represents whether register $R_j$ is zero or positive (there should be precisely
four such 3-tuples, one for each possibility), and $k$ is the label of the next instruction, or  $N$, which signifies halting.
Instructions have to be valid: they never decrement a null register.
A configuration of the machine is written as $(i, r_1,r_2)$ where $i$ is the program counter
and $r_j$ the contents of $R_j$.

It is quite easy to see that the set of possible instructions can be limited without weakening
the model. In particular, the following set of three combinations of $(D_1,D_2)$ suffices:
$(+1,0), (0,-1), (-1,+1)$. 

\subsection*{The Reduction}

Let a 2-counter machine $M$ be given.
The idea is to encode the register values $(r_1,r_2)$ by an $x$ value satisfying
$x = 2^{r_1} 3^{r_2} q$, for some $q$ not divisible by $2,3$. The control state $i$
will be simulated by the $i$th cycle.
Observe that $x \bmod 6$ determines whether each register is null or not.

Each cycle of the mouse represents an instruction, one of the
types $(+1,0), (0,-1)$ and $(-1,+1)$.
They all have $6$ ``essential" states (states moving north and not preceded by an $E$), so that when
the mouse hits the diagonal $y=x$, its state indicates $x \bmod 6$. As explained by van Emde Boas
and Karpinski, the next value of $x$ depends on the previous one as 
$x + \lfloor x/6 \rfloor \cdot b + t$, where $b$ is fixed by the design of the cycle, and $t$ depends on 
the design of the cycle and the value of $x \bmod 6$. 

The cycle $(EN\,EN\,N)^6$ implements the update $(+1,0)$. In fact, it multiplies $x$ by 2.

The cycle  $((EN)^5 \, NNN )^2$  implements the update $(0,-1)$. It multiplies $x$ by $10/6
= 5/3$,
so that $2^{r_1} 3^{r_2} q$ (with $r_2 > 0$) becomes $2^{r_1} 3^{r_2 - 1} (5q)$.

The cycle $((EN)^3 \,NN)^3$ implements the update $(-1,+1)$. In fact, it multiplies $x$ by $9/6
= 3/2$,
so that $2^{r_1} 3^{r_2} q$ (with $r_1 > 0$) becomes $2^{r_1 - 1} 3^{r_2 + 1} q$.

The update of the program counter is implemented in a simple way by the transition function $\delta$
of the super-mouse.

It is possible to restrict super-mice so that all cycles have equal length. This is achieved by choosing
the length as the least common multiple of the lengths of the cycles above (30, 26 and 24): each 
cycle of the above construction is ``pumped up" to the common length by simply repeating the
string several times. It is easy to verify that the mouse can still carry the same computation.

\subsection*{A Variation}

It is also possible to prove that the undecidability already holds for a certain (large enough,
but bounded) number of cycles.
For this purpose, we use a universal CM.  The first cycle will create an initial value of $x$ in such a way
as to ``load" the machine's register with
some encoding of a machine to be simulated. It then jumps into a fixed set of cycles, representing
the part of the machine that does the simulation.

\subsection*{Acknowledgement}

The problem has been brought to my attention by Peter van~Emde Boas during STACS 2013 in Kiel.


\begin{thebibliography}{1}

\bibitem{pultr1982}
A.~Pultr and J.~{\'U}lehla.
\newblock On two problems of mice.
\newblock {\em Proceedings of the 10th Winter School on Abstract Analysis},
  pages 249--262, 1982.

\bibitem{vEB:1980}
P.~van Emde~Boas and M.~Karpinski.
\newblock A number theoretic problem arising from a problem in automata theory.
\newblock {\em Bulletin of the {EATCS}}, 12:50--53, 1980.

\end{thebibliography}
\end{document}